\documentclass[10pt]{article}
\usepackage{bm}
\usepackage{amsmath}
\usepackage{amssymb}
\newcommand{\be}{\begin{equation}}
\newcommand{\ee}{\end{equation}}
\newcommand{\bc}{\begin{center}}
\newcommand{\ec}{\end{center}}
\newcommand{\bea}{\begin{eqnarray}}
\newcommand{\eea}{\end{eqnarray}}
\newcommand{\ben}{\begin{enumerate}}
\newcommand{\een}{\end{enumerate}}
\newcommand{\bean}{\begin{eqnarray*}}
\newcommand{\eean}{\end{eqnarray*}}
\newcommand{\ba}{\begin{array}{l}}
\newcommand{\ea}{\end{array}}
\newcommand{\fc}{\frac}
\newcommand{\bb}{\begin{thebibliography}{99}}
\newcommand{\eb}{\end{thebibliography}}
\newcommand{\ci}[1]{\cite{#1}}
\newcommand{\lab}[1]{\label{#1}}
\newcommand{\re}[1]{(\ref{#1})}

\newcommand{\Ds}{\displaystyle}
\newcommand{\lt}{\left(}
\newcommand{\rt}{\right)}

\newcommand{\cL}{{\cal{L}}}

\newcommand{\pl}{\partial}
\newcommand{\fr}{\frac12}
\newcommand{\gh}[2]{g^{#1 #2}}
\newcommand{\h}[2]{h^{#1 #2}}
\newcommand{\g}[2]{g_{#1 #2}}
\newcommand{\Krs}[3]{\Gamma^{#1}_{ #2 #3}}

\title{\bc  ON THE  HAMILTONIAN FORM OF THE EINSTEIN EQUATIONS FOR GRAVITATIONAL FIELD\ec}
\author{ B.A.FAYZULLAEV\\
Theoretical Physics Department,\\ National University of Uzbekistan, \\
  Tashkent 700095, Uzbekistan}

\begin{document}
\begin{titlepage}
\maketitle

\begin{abstract}
It is shown that Einstein gravitational  equations and canonical equations following from the Dirac-Schwinger Hamiltonian
in the Faddeev variables
coincide.
For proving of this at first, the Einstein equations has been rewritten  in canonical variables,
and after this, the time derivative of the generalized momenta of the gravitational field in Faddeev form has been
calculated using canonical Poisson brackets. The results  coincide.
\end{abstract}
\thispagestyle{empty}
\end{titlepage}

\sloppy
\section{Introduction}

In \cite{D2,schw} the Hamiltonian for the gravitational field was derived (by different methods - in \cite{D2} the constrained dynamics was used, in \cite{schw} by the tetrad formalism).
Detailed consideration  of  this result in the light of the Generalized Hamiltonian Dynamics including the proof of positiveness
of the gravitational energy (in asymptotically flat space-time) based on this hamiltonian can be found in \cite{fad}.
There is another Hamiltonian  formulation of the Einstein gravitation - the Arnowit-Deser-Misner (ADM) formulation \cite{adm}.
In the ADM formalism the variables has been chosen from geometrical viewpoint, mainly.

The Hamilton dynamics for  gravitational field is actively discussing \cite{ghal,kk,kkrv,shest}
including discussions of problems of equations of motion \cite{ay}.
In particular, in \cite{kk,kkrv,shest} a problem of canonical equivalentness of ADM and Dirac formulations is considered.
Authors of \cite{kk} conclude that  Hamiltonians of ADM and Dirac are not related by a canonical transformation. Moreover,
in \cite{kk}  is stated that the ADM variables are not the canonical variables for GR, although they have  simple geometrical meaning.
The derivation of the Einstein's equations by means of the variational principle from the Hilbert's action $S=\int d^4x R$, where the $R$
is the Hilbert's Lagrangian, is one of standard textbook methods.
In this content a question of thorough test of the equivalentness of the Einstein general relativity and its canonical formulations arises.

The general theory of constrained systems \cite{D1,HT,shabproh} states that there are some ambiguities when passing from a Lagrangian to the corresponding Hamiltonian.
That is, there is no strong criterion what exact form a Hamiltonian must take in a constrained theory \cite{shest}.

In this article the question of equivalentness of the Dirac's Gravitational Hamiltonian in the Schwinger-Faddeev variables \cite{fad} with the  Einstein GR is considered.
It is shown that canonical field equations  for pure gravity derived using Poisson brackets
with the Dirac hamiltonian coincides  with the Einstein's gravitational field  equations.

\section{ Hamilton function for gravitational field}
Let's take the following action
$$
S=\int d^{\,4}x\sqrt{-g}\cL=\int dt\int d^{\,3}r \sqrt{-g}\cL,
$$
where the Lagrangian is defined as the Hilbert Lagrangian minus a full divergence:
$$
\sqrt{-g}\cL=\sqrt{-g}R
+\pl_\nu\lt \sqrt{-g}g^{\nu\sigma} \Gamma^\rho_{\sigma\rho}\rt-
\pl_\lambda\lt \sqrt{-g}g^{\nu\sigma} \Gamma^\lambda_{\nu\sigma}\rt,
$$
and is equal to
\be
\sqrt{-g}\cL=h^{\nu\sigma}\left(\Gamma^\lambda_{\rho\lambda}\Gamma^\rho_{\nu\sigma}-\Gamma^\lambda_{\rho\sigma}\Gamma^\rho_{\nu\lambda}\right)+
\pl_\nu h^{\nu\sigma} \Gamma^\rho_{\sigma\rho}-
\pl_\lambda h^{\nu\sigma} \Gamma^\lambda_{\nu\sigma}.
\label{gr12}\ee
Introduced here quantities are follows:
$$R=g^{\mu\nu}R_{\mu\nu},\qquad  g=\det(g_{\mu\nu}),\qquad h^{\mu\nu}=\sqrt{-g}g^{\mu\nu}.$$
and $$R_{\mu\nu}=\partial_\lambda\Gamma^\lambda_{\mu\nu}-\partial_\nu\Gamma^\rho_{\mu\rho}+\Gamma^\sigma_{\mu\nu}\Gamma^\rho_{\sigma\rho} -\Gamma^\rho_{\mu\sigma}\Gamma^\sigma_{\nu\rho}.$$

The detailed discussion of these variables
may be found in \ci{fad}. The Lagrangian $\sqrt{-g}\cL$ corresponds to the first order formalism where $h^{\mu\nu}$ and $\Gamma^\lambda_{\sigma\rho}$
are considered as independent variables.
Equations of motion for \re{gr12} in this case are
$$
\pl_\mu\fc{\pl\sqrt{-g} \cL}{\pl h^{\nu\lambda}_{\quad,\,\,\mu}}-\fc{\pl\sqrt{-g}\cL}{\pl h^{\nu\lambda}}=0,\quad
\pl_\mu\fc{\pl\sqrt{-g} \cL}{\pl {\Gamma^\sigma_{\nu\lambda}}_{,\,\mu}}-\fc{\pl\sqrt{-g}\cL}{\pl\Gamma^\sigma_{\nu\lambda}}=0
$$
which leads to
\be
\Ds{\Gamma^\lambda_{\rho\lambda}\Gamma^\rho_{\nu\sigma}-\Gamma^\lambda_{\rho\sigma}\Gamma^\rho_{\nu\lambda} =\pl_\nu\Krs\rho\sigma\rho-\pl_\rho\Krs\rho\nu\sigma; }
\lab{rmunu2}\ee
\be
\Ds{-\pl_\sigma h^{\nu\lambda}+\delta^\lambda_\sigma\pl_\mu h^{\mu\nu}+ h^{\nu\lambda}\Gamma^\rho_{\sigma\rho}+h^{\rho\mu}\Gamma^\nu_{\rho\mu}\delta^\lambda_\sigma=h^{\nu\mu}\Gamma^\lambda_{\sigma\mu}+h^{\mu\lambda}\Gamma^\nu_{\mu\sigma}. }
\lab{hmunu}\ee
The first of these equations is equivalent to Einstein's pure gravitational field equations:
\be R_{\nu\sigma}=0. \label{rmunu}\ee
The Eq.(\ref{hmunu}) we can rewrite in the form:
\be
\pl_{\sigma}h^{\nu\lambda}=h^{\nu\lambda}\Gamma_{\sigma\rho}^{\rho}-h^{\nu\mu}\Gamma_{\sigma\mu}^{\lambda}-h^{\mu\lambda}\Gamma_{\mu\sigma}^{\nu}
\lab{hgg}\ee
which is nothing but the condition of covariant constantness of the $h^{\nu\lambda}$ (taking into account that it is a tensor density of weight -1):
\be
\nabla_{\sigma}h^{\nu\lambda}=0.
\ee
It turns out to be \cite{D2,schw,fad} that following quantities are generalized coordinates and generalized momenta, respectively, of the gravitational field:
\be
q^{ik}=\h{0}{i}\h{0}{k}-\h{0}{0}\h{i}{k},\qquad\quad \Pi_{ij}=\fc{1}{\h{0}0}\Krs{0}i{j}.
\lab{qpi0}\ee
As it is shown in \cite{schw,fad} our lagrangian may be brought to the form:
\be\ba
\Ds{\sqrt{-g}\cL=\Pi_{ij}\partial_{0}q^{ij} -\frac{1}{h^{00}}q^{ij}q^{kl}\left(\Pi_{ij}\Pi_{kl}-\Pi_{ik}\Pi_{jl}\right)+\partial_{i}\partial_{j}q^{ij}-}\\ \\
\Ds{-2\frac{h^{0l}}{h^{00}}\nabla_{l}\left(q^{j\,k}\Pi_{j\,k}\right)+2\frac{h^{0i}}{h^{00}}\nabla_{j}\left(q^{j\,k}\Pi_{k\,i}\right)
-\frac{1}{h^{00}}q^{ij}R^{(3)}_{ij}.  } \ea\lab{lh}\ee
Here $\nabla_i$ is covariant derivative with respect to the three-space metric (see Appendix A).
From
Eq.\re{lh} we see, that $\Pi_{ij}$ and $q^{ij}$ are canonical
conjugated variables - the $\Pi_{ij}$ are generalized momenta (the tensor density of the weight -1) and
the $q^{ij}$ are generalized coordinates (the tensor density of the weight 2), respectively. But except them we see other variables too - $\Ds{\frac{1}{h^{00}}}$ and $\Ds{\frac{h^{0i}}{h^{00}}}$.
There are no conjugated momenta or velocities for them, this means they are Lagrange multipliers. Variing over them we can obtain
constraints:
\be
C_0=q^{ij}q^{kl} \left(\Pi_{ij}\Pi_{kl}-\Pi_{il}\Pi_{kj} \right)+\gamma R^{(3)}=0,
\ee
where
\be
\gamma R^{(3)}=q^{ij}R^{(3)}_{ij},
\ee
and
\be
C_i=2\nabla_{i}\left(q^{jk}\Pi_{jk}\right)-2\nabla_{j}\left(q^{jk}\Pi_{ki}\right)=0.
\ee
It is more convenient to take as Lagrange multiplier not $\Ds{\frac{1}{h^{00}}}$ but $\Ds{\lambda_0=1+\frac{1}{h^{00}}}.$

Resulting expression for our lagrangian brought to the canonical form is:
\be
L=\int d^{\,3}x \left(\Pi_{ij}\partial_0 q^{ij}-\lambda^{0}C_0-\lambda^{i}C_i-\cal{H} \right).
\lab{lagham}\ee
Hamilton function is
$$
{\cal{H}}=-C_0-\partial_i\partial_j q^{ij}.
$$
Lagrange multipliers here are chosen in the form:
$$
\lambda^0=1+\frac{1}{h^{00}},\qquad\lambda^i=\frac{h^{0i}}{h^{00}}.
$$
As it was shown by Dirac \ci{D1} as time displacement generator we should consider
the so-called total Hamiltonian. In our case it is
\be\ba
\Ds{{\cal{H}}_T={\cal{H}}+\lambda^{0}C_0+\lambda^{i}C_i=\frac{1}{h^{00}}q^{ij}q^{kl}\left(\Pi_{ij}\Pi_{kl}-\Pi_{ik}\Pi_{jl}\right)-\partial_{i}\partial_{j}q^{ij}+}\\ \\
\Ds{+2\frac{h^{0l}}{h^{00}}\nabla_{l}\left(q^{j\,k}\Pi_{j\,k}\right)-2\frac{h^{0i}}{h^{00}}\nabla_{j}\left(q^{j\,k}\Pi_{k\,i}\right)
+\frac{1}{h^{00}}q^{ij}R^{(3)}_{ij} .}
\ea\label{htotal}\ee
A $R^{(3)}_{ij}$ appeared here is 3-dimensional space Ricci tensor (see the Appendix \ref{sdwn}) and  $R^{(3)}$ - is its Ricci scalar.
On the surface of constraints we can go only after
all the Poisson brackets are calculated.

\section{Einstein's equations for gravitational field}
For pure gravitational field Einstein's equations  are
\be
R_{\mu\nu}=0.
\ee
Of them $R_{00}=0$ and $R_{0i}=0$ not contain second time derivatives.
It is well known that equations containing second time derivatives  are
\be
R_{ij}=\partial_{\lambda}\Gamma^{\lambda}_{ij}-\partial_{j}\Gamma^{\lambda}_{i\lambda}+\Gamma^{\sigma}_{ij}\Gamma^{\rho}_{\sigma\rho}-
\Gamma^{\rho}_{j\sigma}\Gamma^{\sigma}_{i\rho}=0
\label{trij}\ee
For comparison of these equations
with canonical  ones we need to transform it to canonical  variables.
For this it is sufficient  to use
 formulas in Sec.(\ref{sdwn}).

 Let's begin with the first term in Eq.(\ref{trij}):
\be
\partial_{\lambda}\Gamma^{\lambda}_{ij}=\partial_{0}(h^{00}\Pi_{ij})+\partial_{k}\gamma^{k}_{ij}+\partial_{k}(h^{0k}\Pi_{ij}),\\ \\
\ee
Using
 \re{hgg} with zero indices:
\be
\partial_{0}h^{00}=-h^{00}\Gamma^{0}_{00}+h^{00}\Gamma^{l}_{0l}-2h^{0l}\Gamma^{}_{0l}
\ee
and Eq.(\ref{k0i}) we find
\be
\partial_{\lambda}\Gamma^{\lambda}_{ij}=h^{00}\partial_0\Pi_{ij}+h^{0k}\partial_k\Pi_{ij}+\partial_k\gamma^k_{ij}-\Pi_{ij}h^{00}\Gamma^0_{00}+(h^{0k}h^{0l}-h^{00}h^{kl})\Pi_{kl}\Pi_{kl}+2\frac{h^{0k}}{h^{00}}\nabla_kh^{00}\Pi_{ij}
\ee
The second term takes the form:
\be\Ds{-\partial_{j}\Gamma^{\lambda}_{i\lambda}=\partial_{j}\frac{\nabla_{i}h^{00}}{h^{00}}-\partial_{j}\gamma^{k}_{ik}.}
\ee
The third and fourth terms in the Eq.(\ref{trij}) are cumbersome, although one need only to use Eqs.(\ref{covdiv}),  (\ref{qpi}) and (\ref{k0i}) for substitutions.
For example, presenting third term at the first step as follows
\be
\Gamma^{\sigma}_{ij}\Gamma^{\rho}_{\sigma\rho}=h^{00}\Pi_{ij}\Gamma^{0}_{00}+h^{00}\Pi_{ij}\Gamma^{k}_{0k}+\Gamma^{l}_{ij}\left(\gamma^{k}_{lk}-\frac{\nabla_{l}h^{00}}{h^{00}}\right)
\ee
we see cancellation of expressions with $\Gamma^{0}_{00}$ in the first and third terms.
After all substitutions and simplifications one can find that
\be\ba
\Ds{R_{ij}=h^{00}\partial_0\Pi_{ij}+R^{(3)}_{ij}+2q^{kl}\left(\Pi_{mn}\Pi_{kl}-\Pi_{mk}\Pi_{nl}  \right)+ \Pi_{ij}\left(-\nabla_kh^{0k}+h^{0k}\gamma^l_{kl}+h^{0k}\frac{\nabla_kh^{00}}{h^{00}}  \right) + }\\ \\
\Ds{+ \Pi_{ik}\left(\nabla_jh^{0k}-h^{0l}\gamma^k_{j\,l}-h^{0k}\frac{\nabla_jh^{00}}{h^{00}}  \right)+\Pi_{kj}\left(\nabla_ih^{0k}-h^{0l}\gamma^k_{il}-h^{0k}\frac{\nabla_ih^{00}}{h^{00}}  \right)-  }\\ \\
\Ds{ -\frac{1}{h^{00}}\,\gamma_{ij}^k\nabla_kh^{00}-\frac{\nabla_ih^{00}}{h^{00}}\frac{\nabla_jh^{00}}{h^{00}} +  \partial_j\left(\frac{\nabla_ih^{00}}{h^{00}} \right).}
\ea\ee
We may rewrite it in the fully three-covariant form as follows:
\be\ba
\Ds{\partial_0{\Pi}_{ij}=-\frac{2}{h^{00}}\,q^{kl}\left(\Pi_{ij}\Pi_{kl}-\Pi_{ik}\Pi_{jl}\right)-\frac{1}{h^{00}}R^{(3)}_{ij}+\Pi_{ij}\nabla_k\left(\frac{h^{0k}}{h^{00}}\right)-\Pi_{ik}\nabla_j\left(\frac{h^{0k}}{h^{00}}\right)-}\\ \\
\Ds{-\Pi_{jk}\nabla_i\left(\frac{h^{0k}}{h^{00}}\right)-\frac{h^{0k}}{h^{00}}}\nabla_k\Pi_{ij}+\frac{1}{h^{00}}\frac{\nabla_{i}h^{00}}{h^{00}}\frac{\nabla_{j}h^{00}}{h^{00}}-\frac{1}{h^{00}}\nabla_{j}\left(\frac{\nabla_{i}h^{00}}{h^{00}}\right).
\ea\label{eineq}\ee

\section{Canonical equations for gravitational field}
According to Dirac the total Hamiltonian Eq.(\ref{htotal})
is generator of time displacement. For providing calculations we need in
 the fundamental Poisson bracket which follows from Eq.(\ref{lagham}):
\be
\Bigg\{\Pi_{ik}(t,\textbf{x}), \,q^{\,j\,l}(t,\textbf{y})\Bigg\}=\frac{1}{2}\left(\delta^{j}_{i}\delta^{l}_{k}+\delta^{j}_{k}\delta^{l}_{i}\right)\delta^{(3)}(\textbf{x}-\textbf{y}).
\label{funpb}\ee
Now we can derive equations of motion for gravitational field using our total Hamilton function Eq.\re{htotal}:
\be
\partial_0{\Pi}_{ij}(x)=\int d^{\,3}y\,\Bigg\{{\cal{H}}_T(y), \Pi_{ij}(x)\Bigg\}.
\label{pbham}\ee
It is convenient to divide this Poisson bracket into four  parts:

\be  \int d^3y\Big\{\partial_k\partial_l q^{kl}(y)    ,\,\Pi_{ij}(x)\Big\} ; \label{pb01}\ee
\be  \int d^3y\frac{1}{h^{00}(y)}\Big\{q^{mn}q^{ij}\left(\Pi_{mn}\Pi_{kl}-\Pi_{mk}\Pi_{nl}  \right)(y)    ,\,\Pi_{ij}(x)\Big\}; \label{pb02} \ee
\be  \int d^3y\frac{1}{h^{00}(y)}\Big\{q^{kl}R^{(3)}_{kl}(y)    ,\,\Pi_{ij}(x)\Big\} ; \label{pb03}\ee
\be  \int d^3y\frac{2h^{0l}(y)}{h^{00}(y)}\Big\{  \nabla_{l}\left(q^{p\,k}\Pi_{p\,k}\right)-\nabla_{k}\left(q^{k\,p}\Pi_{p\,l}\right)(y)  ,\,\Pi_{ij}(x)\Big\} \label{pb04}\ee
During the  calculations below we will suppose $x$ means $ (t,\mathbf{x})$ and $y$ means  $ (t,\mathbf{y})$ in arguments of our functions.

It is obvious that
\be
\int d^3y\Big\{\partial_k\partial_l q^{kl}(y)    ,\,\Pi_{ij}(x)\Big\}=0.
\ee
It is easy to calculate the second Poisson bracket Eq.(\ref{pb02}) too:
\be
\int d^3y\frac{1}{h^{00}(y)}\Big\{q^{mn}q^{ij}\left(\Pi_{mn}\Pi_{kl}-\Pi_{mk}\Pi_{nl}  \right)(y)    ,\,\Pi_{ij}(x)\Big\}=-\frac{2}{h^{00}}q^{kl}\left(\Pi_{ij}\Pi_{kl}-\Pi_{ik}\Pi_{jl}  \right).
\ee
The calculation of the Poisson bracket Eq.(\ref{pb03}) containing  $R^{(3)}_{ij}$ is much complicated. At first step we have:
\be
\int d^3y\frac{1}{h^{00}(y)}\Big\{q^{kl}(y)R^{(3)}_{kl}(y), \Pi_{ij}(x)\Big\}=-\frac{1}{h^{00}}R^{(3)}_{ij}+\int d^3y\frac{q^{kl}(y)}{h^{00}(y)}\Big\{R^{(3)}_{kl}(y), \,\Pi_{ij}(x)\Big\}.
\ee
The second term in the r.h.s. may be expressed as follows:
\be\begin{array}{l}
\Ds{\int d^3y\frac{q^{kl}(y)}{h^{00}(y)}\Big\{R^{(3)}_{kl}(y), \,\Pi_{ij}(x)\Big\} = }\\ \\
\Ds{     = -\int d^3y\frac{q^{kl}(y)}{h^{00}(y)}\frac{\delta}{\delta q^{ij}(x)}\Big[\partial_j \gamma^j_{kl}(y)-\partial_l\gamma^j_{j\,k}(y)+\gamma^s_{k\,l}(y)\gamma^j_{j\,s}(y)-\gamma^s_{k\,j}(y)\gamma^j_{l\,s}(y)\Big]  =}\\ \\
\Ds{=\int d^3y\frac{q^{kl}(y)}{h^{00}(y)}\frac{\delta}{\delta q^{ij}(x)}\Big[-\partial_j \gamma^j_{kl}-\frac12 \partial_k\partial_l\ln\sqrt{q}+\frac12\gamma^s_{k\,l}\partial_s\ln\sqrt{q}+\gamma^s_{k\,j}\gamma^j_{l\,s}\Big].}
\end{array}\label{pb1}\ee
We will dwell  only on some steps  of  the calculation.
  For example,  using the following formula
   \be
\int d^3y\frac{q^{kl}}{h^{00}}\frac{\delta}{\delta q^{ij}(x)}\gamma^s_{k\,j}\gamma^j_{l\,s}=-\int d^3y\frac{1}{h^{00}}\left( \partial_sq^{l\,k}+ q^{l\,k}\partial_s\ln\sqrt{q}\right)\frac{\delta}{\delta q^{ij}(x)}\gamma^s_{l\,k},
   \ee
   we see that the sum of the first and fourth terms in the last line of the Eq.(\ref{pb1}) is
   \be
   -\int d^3y \frac{1}{h^{00}} \left( \partial_s\ln\sqrt{q}+\partial_s\right)\left(q^{kl}\frac{\delta}{\delta q^{ij}(x)}\gamma^s_{l\,k} \right),
   \ee
   where the partial derivatives in the first brackets are acting on the $y$ argument, the dependence on $x$ is distinguished.
 The expression in the second bracket more convenient to calculate using the following identity:
 \be
q^{kl}\frac{\delta}{\delta q^{ij}(x)}\gamma^s_{l\,k}=\frac{\delta}{\delta q^{ij}(x)}\left(q^{kl}\gamma^s_{l\,k}\right)-\gamma^s_{ij}\delta(\mathbf{x}-\mathbf{y})
 \ee
 and then to use (\ref{qgamma}).

The final expression for the Poisson bracket is
\be\ba
\Ds{\int d^3y\frac{q^{kl}(y)}{h^{00}(y)}\Big\{R^{(3)}_{kl}(y), \Pi_{ij}(x)\Big\} =\frac{1}{2h^{00}}\gamma_{ij}^k \partial_k\ln\sqrt{q}-\gamma_{ij}^k \partial_k\frac{1}{h^{00}}+\partial_i\partial_j\frac{1}{h^{00}}\,+       }\\ \\
\Ds{ +\frac{1}{4h^{00}}\,\partial_i\ln\sqrt{q}\,\partial_j\ln\sqrt{q}- \frac{1}{2h^{00}}\,\partial_i\partial_j\ln\sqrt{q}-\frac12\left(\partial_i\ln\sqrt{q}\,\partial_j\frac{1}{h^{00}}+\partial_j\ln\sqrt{q}\,\partial_i\frac{1}{h^{00}} \right) .     }
\ea\label{pb32}\ee
Taking into account
\be
\frac{1}{h^{00}}\frac{\nabla_kh^{00}}{h^{00}}=\frac{1}{2h^{00}}\,\partial_k\ln\sqrt{q}-\partial_k\frac{1}{h^{00}}
\ee
 the r.h.s. of the Eq.(\ref{pb32}) has been brought to the following covariant form:
\be\ba
\Ds{\frac{1}{(h^{00})^2}\,\gamma_{ij}^k\nabla_kh^{00}+\frac{1}{h^{00}}\frac{\nabla_ih^{00}}{h^{00}}\frac{\nabla_jh^{00}}{h^{00}} -   \frac{1}{h^{00}}\,\partial_j\left(\frac{\nabla_ih^{00}}{h^{00}} \right)=}\\ \\
\Ds{=\frac{1}{h^{00}}\frac{\nabla_ih^{00}}{h^{00}}\frac{\nabla_jh^{00}}{h^{00}}-   \frac{1}{h^{00}}\,\nabla_j\left(\frac{\nabla_ih^{00}}{h^{00}} \right).}
\ea\ee

Using this result we have for the Poisson bracket Eq.(\ref{pb03}):
\be
\int d^3y\frac{1}{h^{00}(y)}\Big\{q^{kl}(y)R^{(3)}_{kl}(y), \Pi_{ij}(x)\Big\}=-\frac{1}{h^{00}}R^{(3)}_{ij}+\frac{1}{h^{00}}\frac{\nabla_ih^{00}}{h^{00}}\frac{\nabla_jh^{00}}{h^{00}}-   \frac{1}{h^{00}}\,\nabla_j\left(\frac{\nabla_ih^{00}}{h^{00}} \right).
\ee

The last bracket Eq.(\ref{pb04}) is
\be\ba
\Ds{\int d^3y\frac{2h^{0l}(y)}{h^{00}(y)}\Big\{  \nabla_{l}\left(q^{p\,k}\Pi_{p\,k}\right)-\nabla_{k}\left(q^{k\,p}\Pi_{p\,l}\right)(y)  ,\,\Pi_{ij}(x)\Big\}   =}\\ \\
\Ds{=\int d^3y\frac{2h^{0l}(y)}{h^{00}(y)}\Big\{\partial_{l}\left(q^{p\,k}\Pi_{p\,k}\right)-\partial_{k}\left(q^{k\,p}\Pi_{p\,l}\right)-\frac12 \partial_lq^{ks}\Pi_{ks}  ,\,\Pi_{ij}(x)\Big\} = }\\ \\
\Ds{=2\Pi_{ij}\partial_l\left(\frac{h^{0l}}{h^{00}}\right) -\Pi_{jl}\partial_i\left(\frac{h^{0l}}{h^{00}}\right)-\Pi_{il}\partial_j\left(\frac{h^{0l}}{h^{00}}\right)-\partial_l\left(\Pi_{ij}\frac{h^{0l}}{h^{00}}\right)=   }\\ \\
\Ds{=\Pi_{ij}\nabla_l\left(\frac{h^{0l}}{h^{00}}\right)-\frac{h^{0l}}{h^{00}}\nabla_l\Pi_{ij}- \Pi_{jl}\nabla_i\left(\frac{h^{0l}}{h^{00}}\right)-\Pi_{il}\nabla_j\left(\frac{h^{0l}}{h^{00}}\right)  .  }
\ea\ee
Summing up all contributions we obtain:
\be\ba
\Ds{\partial_0{\Pi}_{ij}=-\frac{2}{h^{00}}\,q^{kl}\left(\Pi_{ij}\Pi_{kl}-\Pi_{ik}\Pi_{jl}\right)-\frac{1}{h^{00}}R^{(3)}_{ij}+\Pi_{ij}\nabla_k\left(\frac{h^{0k}}{h^{00}}\right)-\Pi_{ik}\nabla_j\left(\frac{h^{0k}}{h^{00}}\right)-}\\ \\
\Ds{-\Pi_{jk}\nabla_i\left(\frac{h^{0k}}{h^{00}}\right)-\frac{h^{0k}}{h^{00}}}\nabla_k\Pi_{ij}+\frac{1}{h^{00}}\frac{\nabla_{i}h^{00}}{h^{00}}\frac{\nabla_{j}h^{00}}{h^{00}}-\frac{1}{h^{00}}\nabla_{j}\left(\frac{\nabla_{i}h^{00}}{h^{00}}\right).
\ea\label{caneq}\ee
The result is the same as the Eq.(\ref{eineq}).
Thus Einstein equations of motion for gravitational field and canonical equations obtained using Hamiltonian Eq.(\ref{htotal}) coincide.

\appendix

\section{Definitions of 3-dimensional variables\label{sdwn}}

In Hamiltonian dynamics time and space coordinates enter separately, for this we must introduce some corresponding 3-space notations.
As we have mentioned in the Introduction we will use Faddeev's variables \cite{fad}.
For example, a contravariant 3-space metric tensor is defined as follows:
$$
\gamma^{ij}g_{j\,k}=\delta^i_k,
$$
from which one find:
\be
\Ds{\gamma^{ij}=g^{i\,j}-\fc{\gh{0}i\gh{0}j}{\gh{0}0}}.
\lab{gamij}\ee
Accordingly, for Christoffel symbols in 3-space we have:
\be
\gamma^i_{jk}=\fr \gamma^{ik}\lt \pl_j\g{i}k+\pl_k\g{l}j-\pl_l\g{j}k \rt.
\label{gammasm}\ee
Three-space Ricci tensor is defined as
\be
R_{ij}^{(3)}=\partial_{l}\gamma^{l}_{ij}-\partial_{j}\gamma^{l}_{li}+\gamma^{l}_{lk}\gamma^{k}_{ij}-\gamma^{l}_{ik}\gamma^{k}_{lj}.
\lab{r3ij}\ee
Denoting
\be
\gamma=\det (g_{ij})\quad {\rm{and}}\quad
g=\det{g_{\mu\nu}}
\ee
we conclude that
\be
g^{00}=\gamma/g.
\ee
It is not difficult to see that
\be
q^{ij}=\gamma\gamma^{ij},\qquad q_{ij}=\frac{1}{\gamma}g_{ij},\qquad q=\det{q_{ij}}=\frac{1}{\gamma^2}.
\ee
Our variables $h^{\nu\lambda}$ are both
4-space and 3-space densities with weight (-1), so
 we have for their 3-space covariant derivatives:
\be
\nabla_i\h{0}0=\pl_i\h{0}0-\gamma^j_{ij}\h{0}0,\quad
\nabla_i\h{0}k=\pl_i\h{0}k+\gamma^k_{il}\h{0}l-\gamma^j_{ij}\h{0}k.
\lab{covdiv}\ee
But quantities like $h^{0i}/h^{00}$ are 3-vectors (not densities) and, for example, we have
\be
\Big[\nabla_{j},\nabla_{k}\Big]\frac{h^{0k}}{h^{00}}=-R^{(3)}_{j\,k}\frac{h^{0k}}{h^{00}}.
\lab{rijh}\ee
Following quantities are generalized coordinates and generalized momenta of gravitational field, respectively:
\be
q^{ik}=\h{0}{i}\h{0}{k}-\h{0}{0}\h{i}{k},\qquad\quad \Pi_{ij}=\fc{1}{\h{0}0}\Krs{0}i{j}.
\lab{qpi}\ee
Let's to write out that part of Eq.\re{hgg} which contains no time derivatives:
\be
\pl_i h^{kj}-h^{kj}\Gamma^\rho_{i\rho}+h^{\mu k}\Gamma^{j}_{i\mu}+h^{\mu j}\Gamma^{k}_{\mu i}=0;
\lab{hij}\ee
\be
\pl_i\h{0}{0}+\h{0}{0}\Krs{0}{0}{i}+2\h{0}{j}\Krs{0}{i}{j}-\h{0}{0}\Krs{j}{i}{j}=0;
\lab{h00}\ee
\be
\pl_i\h{0}{j}+\h{0}{0}\Krs{j}{i}0+\h{0}{k}\Krs{j}{i}{k}+\h{k}{j}\Krs{0}k{i}-\h{0}{j}\Krs{k}i{k}=0.
\lab{h0i}\ee
Using  these equations and definitions of the generalized coordinates and momenta
we can  express some Christoffel symbols we need as follows:
\be
\Krs{k}{i}0=-\fc{1}{\h{0}0}\nabla_i\h{0}k-\h{l}k \Pi_{il},\quad
\Krs{0}{i}0=-\fc{1}{\h{0}0}\nabla_i\h{0}0-\h{j}0 \Pi_{ji},\quad \Krs{i}{j}k=\gamma^i_{jk}+{\h{0}i}\Pi_{j\,k}.
\lab{k0i}\ee
Contributions of all Christoffel symbols of the type $\Gamma^i_{00}$ are canceling. Three dimensional Christoffel symbols are defined in Eq.(\ref{gammasm}).
For the calculation of Poisson brackets it is necessary  the form of  the $\gamma^i_{jk}$  when it depends only on canonical variables $q^{ij}=\gamma \gamma^{ij}$ and $\Ds{q_{ij}=\frac{1}{\gamma}g_{ij}}$:
\be
\gamma^i_{j\,k}=\frac12 q^{i\,l}\left(\partial_jq_{kl}+\partial_kq_{j\,l}-\partial_lq_{jk}\right)+\frac12 \left(q^{i\,l}q_{j\,k}\partial_l-\left(\delta^i_j\partial_k+\delta^i_k\partial_j\right)\right)\ln\sqrt{q}.
\ee
We have also
\be
\gamma^k_{i\,k}=\partial_i\ln\sqrt{\gamma}=-\frac12\partial_i\ln\sqrt{q}
\ee
and
\be
q^{j\,k}\gamma^i_{j\,k}=-\partial_kq^{i\,k}-\frac12\, q^{i\,k}\partial_k\ln\sqrt{q}.
\label{qgamma}\ee

\bb

\bibitem{D2} Dirac P.A.M., The theory of gravitation in Hamiltonian form, Proc.Roy.Soc.,  {\bf A246}(1958)333.
\bibitem{schw} Schwinger J., Quantized gravitational field, Phys.Rev., \textbf{130}(1963)1253.
\bibitem{fad} Faddeev L.D., The problem of energy in Einstein's theory of gravitation,  Sov.Phys.Usp., \textbf{25}(1982)130 [Usp.Fiz.Nauk, \textbf{136}(1982)435]
\bibitem{adm} Arnowitt R., Deser S. and Misner C.W., The dynamics of General Relativity, in \textit{Gravitation: an introduction to current research}, ed. by L.Witten, pp.227-264, (Wiley, NY 1962)[gr-qc/0405109]
\bibitem{ghal} Ghalati R.N., A novel Hamilton formulation of first order Einstein-Hilbert action: connection with ADM, diffeomorphism invariance and linearized theory, gr-qc/0901.3344
\bibitem{kk} Kiriushcheva N. and S.V.Kuzmin, The Hamiltonian formulation of General Relativity: myths and reality, gr-qc/0809.0097
\bibitem{kkrv} Kiriushcheva N., S.V.Kuzmin, C.Racknor and S.R.Valluri, Diffeomorphism invariance in the Hamiltonian formulation of General Relativity, Phys.Lett.A372(2008)5101[gr-qc/0808.2623]
\bibitem{shest} Shestakova T.P., Hamilton formulation of General Relativity 50 years after the Dirac celebrated paper: do unsolved problems still exist? [gr-qc/0911.5252]
\bibitem{ay} Anderson A. and J.W.York, Hamiltonian time evolution for General relativity, Phys.Rev.Lett.\textbf{81}(1998)1151-1157[gr-qc/9807041]
\bibitem{D1} Dirac P.A.M., Lectures on Quantum Mechanics, Yeshiva University, NY,(1964)
\bibitem{HT} Henneaux M. and Teitelboim C. Quantization of Gauge Systems, Princeton Univ.Press., Princeton, NJ(1994)
\bibitem{shabproh} Prokhorov L.V. and Shabanov S.V., Hamiltonian mechanics of gauge systems (in Russian), St.Petersbourg Univ.Press, (1997)
\eb

\end{document}